      [1999/12/01 v1.4c Il Nuovo Cimento]
\documentclass{cimento}

\title{Fourth Generation --- Towards Effect of Large Yukawa Coupling
}
\author{George W.S. Hou\from{ins:NTU}\from{ins:NCTSn}\thanks{e-mail: wshou@phys.ntu.edu.tw}}
\instlist{\inst{ins:NTU} Department of Physics, National Taiwan
    University -- Taipei, Taiwan
  \inst{ins:NCTSn} NCTS-North,
    National Taiwan University -- Taipei, Taiwan}
\PACSes{\PACSit{}{\ 11.30.Er,\ 11.30.Hv,\ 12.60.Jv,\ 13.25.Hw}}
\begin{document}

\maketitle

\begin{abstract}
In this talk we cover two threads regarding the 4th generation:
 on CP violation, from Earth up to the Heavens,
  {\it i.e.} from accelerator-based experimental studies,
  towards baryon asymmetry of the Universe;
 and on direct search bounds on $m_{t'}$ and $m_{b'}$, towards the
  possibility of electroweak symmetry breaking through large Yukawa couplings.
Prospects and discussions are presented.
\end{abstract}

\section{Introduction:  the Four Statements}

Interest in the fourth generation (4G, or SM4) seems warming up.
Not only there is this dedicated invited talk at the TOP2010 workshop, but
the summary ``Four Statements about the Fourth Generation"~\cite{ref:4S4G}
of a dedicated workshop held in September 2008 at CERN,
received 50+ citations within a year of posting.
The four statements are:
\begin{enumerate}
\item The fourth generation is not excluded by EW precision data.
\item SM4 addresses some of the currently open questions.
\item SM4 can accommodate emerging possible hints of new physics.
\item LHC has the potential to discover or fully exclude SM4.
\end{enumerate}
A followup \emph{2nd Workshop on} ``\emph{Beyond 3 Generation Standard
Model}", subtitled ``\emph{New Fermions at the Crossroads of Tevatron
and LHC}",

\centerline{http://indico.cern.ch/conferenceDisplay.py?confId=68036}

\noindent was held January 2010 in Taipei. It would likely be
followed up further.

In this talk, we limit ourselves to two threads related to
the 4th generation. The first thread is on CP violation (CPV).
We start from ``hints from B factories" that link to
``CPV in $B_s \to J/\psi \phi$ at Tevatron" (both under Statement 3),
and discuss ``LHCb prospects" (under Statement 4). The real importance is
beyond CPV studies on Earth, but with Heavenly implications,
{\it i.e.} the possibility to address ``new CPV source for BAU"
(Baryon Asymmetry of the Universe, which is under Statement 2)!

A second thread starts from ``Tevatron direct search" (under
Statement 3) for the $t'$ and $b'$ quarks, which naturally links to
``ATLAS and CMS discovery prospects" (under Statement 4).
This offers ``new perspective on Higgs naturalness" (under Statement2),
in particular, perhaps touching on the mechanism underlying 
electroweak symmetry breaking.
As such, it could ``impact on Higgs searches" at the LHC (under Statement 4).

%\begin{figure}
%\includegraphics{foo}     % includes figure foo.eps
%\caption{Foo onteracting with bar}
%\end{figure}

\section{CPV4U:  from Earth to Heaven}

\subsection{Earthly Thread}

Direct CPV (DCPV) in $B \to K\pi$ decay, and
the difference between charged and neutral modes,
are rather personal to me.

In his plenary talk at ICHEP2004, Yoshi Sakai (now a Belle spokesperson)
showed the emerging 2.4$\sigma$ difference between DCPV in
$B^+\to K^+\pi^0$ and $B^0 \to K^+\pi^-$ from the Belle experiment.
He showed the $Z$-penguin diagram for $B^- \to K^-\pi^0$ decay,
and offered the questions: Large EW penguin? New Physics?
These words were coming from my first draft writing of Ref.~\cite{ref:BelleAKpi}.
Personally, I was quite shaken. Though not yet significant (but
consistent with BaBar!), the difference is staggeringly large,
larger than the $-10\%$ asymmetry in the neutral $B$ mode.
The latter is generated by the strong penguin, with the help of some
``hadronic" $CP$ conserving phase. How can some effect that enters
this difference between the $B^+\to K^+\pi^0$ and $B^0 \to K^+\pi^-$ modes be so large?

I recalled my first B paper --- also my first 4G paper ---
on the nondecoupling of large top quark mass effect on $b\to
s\ell^+\ell^-$ (and $s\nu\bar\nu$) rate~\cite{ref:btosll}.
Compared with the photonic penguin, where naive counting gives it at
$\alpha G_F$ order, one would have thrown away the $Z$-penguin,
which is at $G_F^2$ order. Even after noticing the
mismatch of mass dimensions, taking $m$ at $m_b$ scale,
$G_F m_b^2$ is still much smaller than $\alpha$.
Direct computation, however, showed that $m$ is closer to $m_t$,
and the $Z$-penguin can overwhelm the photonic penguin
for $m_t \sim M_W$ or heavier. This evasion of the familiar
decoupling theorem (which works in QED and QCD) is because the
heavy mass not only appears in propagators (hence damps the
amplitude --- decoupling), in spontaneously broken chiral gauge
theories, masses can appear in the numerator as Yukawa couplings.
Thus, this nondecoupling is a special dynamical phenomena in
$\chi$GT that undergo SSB, such as the electroweak theory.

With this backdrop, I worked out, with Makiko Nagashima and Andrea
Soddu, that~\cite{ref:HNS} the $t'$ quark could indeed jack up
$A_{K^+\pi^0}$, turning it positive as compared with
$A_{K^+\pi^-} \simeq -10\%$.
Besides enjoying nondecoupling, the heavy $t'$ quark brings in
new CPV phase in $V_{t's}^*V_{t'b}$, while $V_{ts}^*V_{tb}$ has
practically no phase within 3G.

The nondecoupling of $t$ and $t'$ in the $Z$-penguin is echoed in
the better known box diagram for $B^0$--$\bar B^0$ mixing.
%(a box diagram in fact accompanies the $Z$-penguin by gauge invariance).
Thus, if there is a large CPV effect in the $b\to s$ $Z$-penguin,
a corollary is that the box diagram for $B^0_s$--$\bar B^0_s$ mixing
would pick up a large CPV phase, and tagged time-dependent CPV
(TCPV) in $B_s \to J/\psi\,\phi$, as well as $A_{\rm SL}$, should yielded
a large effect. A prediction~\cite{ref:HNS} of $\sin2\Phi_{B_s}
\sim -0.2$ to $-0.7$ was given for TCPV in $B_s \to J/\psi\,\phi$.
When $\Delta m_{B_s}$ became precisely measured by CDF in 2006, we
refined the prediction~\cite{ref:HNS07} to
\begin{equation}
 \label{e.sin2PhiBs}
 \sin2\Phi_{B_s} \sim -0.5\ \un{--}\ -0.7, \quad\quad {(m_{t'} = 300\ \rm
 GeV)}
\end{equation}
by comparing $\Delta m_{B_s}$ with ${\cal B}(B \to
s\ell^+\ell^-)$. We used $A_{K\pi}$ difference only to select the
sign of $\sin2\Phi_{B_s}$, since there is some controversy of
``hadronic effect" on the $A_{K\pi}$ difference.

Thus, large CPV in $B_s$ mixing is possible, despite $\Delta
m_{B_s}$ and ${\cal B}(B \to s\ell^+\ell^-)$ both being SM-like
(these have $f_{B_s}$ or form factor dependence, respectively).
Although measurement would be a sure thing once LHCb has data,
Tevatron now has a chance for big discovery, and for this sake I
made a dedicated trip to Fermilab in Spring 2007, to present the
case to a joint B audience from CDF and D$\emptyset$.
Somewhat spectacularly, first in December 2007 by CDF, then
February 2008 from D$\emptyset$, followed by a summer 2008
update\footnote{
 In the summer 2008 public note (CDF/ANAL/BOTTOM/PUBLIC/9458) from
 CDF, I had the special honor of being quoted as
  ``George Hou predicted the presence of a $t^\prime$ quark with
   mass between $\sim$ 300 and 1,000 GeV/c$^2$ in order to explain
   the Belle result and predicted \emph{a priori} the observation
   of a large CP-violating phase in $B^0_s \to J/\psi\,\phi$ decays",
   citing Refs.~\cite{ref:HNS} and~\cite{ref:HNS07}.
}
by CDF, three consecutive measurements at the Tevatron yielded
large central values consistent with Eq.~(\ref{e.sin2PhiBs}),
although the significance was not much more than 2$\sigma$. The
SM3 expectation is $-0.04$.
After much work, the final combined
significance~\cite{ref:CDFDzero}, announced at EPS-HEP2009,
is $2.1\sigma$. But more data is in store at the Tevatron.

Indeed, there is very recent new activity. Two weeks before TOP2010,
D$\emptyset$ reported~\cite{ref:Dzero10} a significant signal for
same-sign dimuons,
\begin{equation}
 \label{e.ASLb}
 A_{\rm sl}^{\rm b} = -0.00957 \pm 0.00251\;{\rm (stat)}
                               \pm 0.00146\;{\rm (syst)},
                       \quad\quad {\rm D}\emptyset,\ 6.1\ {\rm fb}^{-1},
\end{equation}
which is a combination of $A_{\rm sl}^s$ and $A_{\rm sl}^d$, the
respective dimuon asymmetries arising from $B_s^0$ and $B_d^0$. 
The result of Eq.~(\ref{e.ASLb}) is $3.2\sigma$ from
the SM expectation of $-0.00023^{+0.00005}_{-0.00006}$, which is
practically zero compared to the D$\emptyset$ value. The central
value is in fact almost identical to the 2007 result, but
D$\emptyset$ was able to reduce the error by a factor of 2, 
strengthening the discrepancy with SM on the
$\sin\phi_s$--$\Delta\Gamma_s$ plane (note that $\sin2\Phi_{B_s} =
- \sin2\beta_s = \sin\phi_s$, where $-2\beta_s$ is the notation of
CDF, while D$\emptyset$ uses $\phi_s$).

Very shortly after the D$\emptyset$ announcement,
CDF updated~\cite{ref:CDF10} (at FPCP2010, the week before TOP2010)
their tagged $B_s \to J/\psi\,\phi$ study, to a dataset of 5.2 fb$^{-1}$,
almost doubling the data from 2008.
$\Delta m_{B_s}$ is remeasured with the new data, which is impressive.
The value for $\beta_s$ has weakened, to only a variance of 0.8$\sigma$ from SM.
What is intriguing is that, adding the extra data, a wedge is drawn
right up the previous most likely $\beta_s$ value, so the
diminished  $\beta_s$ value is actually the combination of equal
likelihood of near zero, or some value even larger than before. In
other words, it smells like large fluctuations.

What can one make of these new developments?
First, D$\emptyset$ and CDF are consistent with each other, though
$\sin2\Phi_{B_s}$ seems weaker than before.
Second, using the formulas and arguments from
Ref.~\cite{ref:HM07}, $A_{\rm sl}^{\rm s}$, a derived quantity
from the measured $A_{\rm sl}^{\rm b}$
(called $A_{\rm SL}^{\rm TeV}$ in Ref.~\cite{ref:HM07}) by
input of $B_s^0$ and $B_d^0$ production fractions,
is equal to $\Delta\Gamma_s^{\rm SM}/\Delta m_s \times \sin\phi_s$.
Hence, $\vert A_{\rm sl}^{\rm s} \vert < 0.008$ if one uses the
Lenz-Nierste result~\cite{ref:LN07} for $\Delta\Gamma_s^{\rm SM}$,
the CDF measured value for $\Delta m_s \equiv \Delta m_{B_s}$, and
saturating $\vert\sin\phi_s\vert$ by 1. That this bound is already
violated by the D$\emptyset$ result means that
$\Delta\Gamma_s^{\rm SM}$ is larger than the Lenz-Nierste
estimate, which implies ``hadronic" enhancement may be present.
This enhancement seems sizable, if one considers a smaller, rather
than larger, value for $\sin2\Phi_{B_s}$ as indicated by CDF
update. Alternatively, assuming that the D$\emptyset$ result
stays, perhaps New Physics affects $\Gamma_{12}^s$~\cite{ref:DHV},
{\it i.e.} width-mixing, which is usually viewed as more exotic
than affecting mass-mixing. The 4th generation does not affect
$\Gamma_{12}^s$ in any significant way.

In fact, because of rising $m_{t'}$ bounds (as we shall soon turn
to), we have reinvestigated the 4G impact with heavier $m_{t'} =
500$ GeV, and found~\cite{ref:HM10}
\begin{equation}
 \label{e.sin2PhiBs10}
 \sin2\Phi_{B_s} \sim -0.33,
  \quad\quad\quad {(m_{t'} = 500\ \rm GeV)},
\end{equation}
This is in remarkable agreement with the softening of
$\sin2\beta_s$ from CDF update, and rhymes also with the direct
$m_{t'}$ bounds that are now higher than the 300 GeV value used in
Eq.~(\ref{e.sin2PhiBs}). The bad news, for Tevatron at least, is
that such lower $\sin2\Phi_{B_s}$ values probably can never be
``observed" at the Tevatron, and would need LHCb to verify.

LHCb data is eagerly awaited.

\subsection{Heavenly Touch --- Towards BAU}

In his acceptance speech at Nobel 2008 ceremonies, Kobayashi
sensei mentions ``Matter dominance of the Universe seems requiring
new source of CP violation." Let us try to understand the meaning
of this statement.

Kobayashi and Maskawa received the Nobel prize because the B
factories, {\it viz.} the BaBar and Belle collaborations, measured
the CPV phase that verified the nontrivial realization of the CKM
unitarity condition $V_{ud}V_{ub}^* + V_{cd}V_{cb}^* +
V_{td}V_{tb}^* = 0$. There is one subtlety, as we all learned in particle
physics class: Any degeneracy of a like-charge quark pair
would allow a freedom to absorb the unique CPV phase within SM3.
Effectively, one goes back to the 2 generation case where of
course there is no CPV. This subtlety is nicely summarized in the
so-called Jarlskog invariant~\cite{ref:Jarlskog} for CPV,
\begin{equation}
 \label{e.J}
 J = (m_t^2 - m_u^2)(m_t^2 - m_c^2)(m_c^2 - m_u^2)
     (m_b^2 - m_d^2)(m_b^2 - m_s^2)(m_s^2 - m_d^2)\,A,
\end{equation}
where $A$ is the area of any unitarity triangle, such as the
aforementioned ``$b \to d$ triangle". $J$ can be
derived~\cite{ref:Jarlskog} from the algebraic quantity ${\rm
Im}\,{\rm det}\,[M_uM_u^\dag,\, M_dM_d^\dag]$. 
What Kobayashi meant, then, is that $J$ is short by 
at least $10^{-10}$ from what is needed for BAU.
To illustrate numerically, compared with the dimensionless number $n_{\bf
B}/n_\gamma$ (baryon over photon density) measured by WMAP to be
$(6.2\pm 0.2) \times 10^{-10}$, a dimensionless analysis 
by normalizing with the phase transition temperature $\sim 100$ GeV,
gives $J \sim 10^{-20}$, hence falling short by over $10^{-10}$.

Let me make a jump and state that Belle published a paper in the
journal {\it Nature} in 2008. In a single paper, Belle measures
both $A_{K^+\pi^0}$ and $A_{K^+\pi^-}$ and finds~\cite{ref:DAKpi}
\begin{equation}
 \label{e.DAKpi}
 \Delta A_{K\pi} \equiv A_{K^+\pi^0} - A_{K^+\pi^-} = 0.164 \pm
 0.037, \quad\quad\quad {\rm Belle, {\it Nature\;2008}}
\end{equation}
which is a 4.4$\sigma$ effect. Note that the effect is stronger
than the measured DCPV in B decay, $A_{K^+\pi^-} \simeq -10\%$.
{\it The difference is large}, experimentally established
(together with BaBar), and \emph{was not predicted}. Across the
Atlantic, however, many dismiss this effect as likely due to
``hadronic effects" in ``enhanced color-suppressed tree" ...

I mention this {\it Nature} paper because, as a
principal author, in trying to ``explain the importance of CPV to
biologists", I literally went ``out of my mind": the mindset was
very different from our daily living as particle
physicist.
One day in early Fall 2007, I noticed that, if one shifts
by one generation, {\it i.e.} from $123 \rightarrow 234$,
Eq.~(\ref{e.J}) becomes
\begin{equation}
 \label{e.J234}
 J_{(2,3,4)}^{sb} = (m_{t'}^2 - m_c^2)(m_{t'}^2 - m_t^2)(m_t^2 - m_c^2)
                    (m_{b'}^2 - m_d^2)(m_{b'}^2 - m_b^2)(m_b^2 - m_s^2)
                   \,A_{234}^{sb},
\end{equation}
where $A_{234}^{sb}$ is an approximate triangle that governs the
large CPV effect in $b\to s$ transitions. Plugging in numbers,
with $m_{t'}$ and $m_{b'}$ ranging in between 300 to 600 GeV,
$J_{(2,3,4)}^{sb}$ (Eq.~(\ref{e.J234})) is typically enhanced by
$10^{13}$ to $10^{15}$ compared to $J$ (Eq.~(\ref{e.J})), and the
gain is mostly through the \emph{large Yukawa couplings} of 4G
quarks. By simply going from 3G to 4G, one seems to gain enough
CPV for BAU! When I filed the simple writeup from a Z\"urich hotel
room in early March 2008, Providence was indicated in the
number~\cite{ref:HouCJP} --- .1234 --- returned by arXiv, for a
paper on the 4th generation! Would Nature use this? One should note that, the
real staggering factor is this 1000 trillion enhancement.
To me, this is the single most important motivation for the
existence of the fourth generation (this line was quoted by David
Shiga in his New Scientist article dated June 1, 2010,
which was stimulated by the D$\emptyset$ result of Eq.~(\ref{e.ASLb})).

It is mind boggling to think that, as we look up to the starry
heavens (Kant!!), what we do on Earth matters --- to understand the
disappearance of antimatter from the Universe! We caution, of
course, that there is still the unresolved issue of the order of
phase transition~\cite{ref:HouCJP}, and it is not clear yet
whether 4G can help resolve it.

\subsection{Unfinished on Earth}

There are other predictions, such as on $A_{\rm FB}(B \to
K^*\ell^+\ell^-)$. But let me give a more general comment on boxes
and penguins.

It was through the quark level box diagram for $K^0$--$\bar K^0$
mixing that we first learned the GIM mechanism, inferred the charm
quark mass, and accounted for $\varepsilon_K$ (with the top).
It was through the $s\to d\bar qq$ electroweak or $Z$-penguin,
that we learned of a diminished $\varepsilon'/\varepsilon$, and of
the strength of $K\to \pi\bar\nu\nu$ (which still awaits precision
measurement).
It was through the quark level box diagram for $B_d^0$--$\bar B_d^0$ mixing
that we first learned that the top is rather heavy,
allowing also the clean measurement of $\sin2\phi_1/\beta$
($\sin2\Phi_{B_d}$ in our definition).
It was through the $Z$-penguin dominance with heavy top that the
$b\to s\ell^+\ell^-$ rate was first estimated in 1986, which was
measured only by 2002. All these effects, practically all the
important FCNC and CPV effects within the 3G KM model (SM3), are due to
effects of nondecoupling, because of large Yukawa coupling of the
top quark (and charm for $\Delta m_K$).

All these were just with 3 generations. If a 4th generation exist,
every aspect above would be touched by $t'$: $B_s$ system,
$A_{\rm FB}$ in $B \to K^*\ell^+\ell^-$ ($B\to X_s\ell^+\ell^-$),
and $K\to \pi\bar\nu\nu$;
or by $b'$, {\it e.g.} in $D^0$--$\bar D^0$ mixing (especially the CPV part).
Besides the study of the $B_s$ system,
of particular importance would be a future measurement of
$K_L \to \pi^0\bar \nu\nu$, which is purely short distance CPV.
It would provide us access to $V_{t'd}$ in the future~\cite{ref:HM10}.
The KOTO experiment (E14) at J-PARC is to be watched.

Of course, nothing beats the direct discovery of the $t'$ and
$b'$ quarks for establishing SM4, and the LHC would soon take
the lead.

\section{Direct Search: Large Yukawa Coupling and EWSB?}

We now turn to the Tevatron thread of direct search for $t'$ and
$b'$ quarks. We would be more cursory, as much work is towards the
future, and no discovery can yet be claimed.

\subsection{Tevatron Thread --- $t'$ and $b'$ Search Status}

CDF has had a long stretch of an effort~\cite{ref:CDFtprime}, lead
by the UC Davis group, in searching for heavy top-like signals.
Specifically, one searches for pair-produced $t'\bar t'$ with $t'
\to Wq$, where one $W$ undergoes leptonic decay, so the signature
is $\ell$ plus missing $p_T$ ($E_T$) plus 4 or more jets. No
$b$-tagging is imposed for sake of efficiency, with the advantage that no
assumptions are made of the decay $q$ flavor from $t'$. Starting from sub-fb$^{-1}$
data, the interesting, if not nagging feature is some high $M_{\rm
reco}$ activity (projected on $M_{\rm reco}$--$H_T$ plane, where
$M_{\rm reco}$ is the reconstructed mass mimicking $m_{t'}$, and
$H_T$ is a variable measuring amount of transverse activity), that
do not seem to go away as the data increases.

In the latest public result in CDF Note
10110~\cite{ref:CDFtprime}, once again the limit on $t'$ cross
section {\it vs.} $m_{t'}$ does not drop as expected,
``saturating" roughly at 0.1 pb$^{-1}$ beyond $\sim 250$ GeV (close
to the published limit~\cite{ref:CDFtprime}), giving the observed
bound
\begin{equation}
 \label{e.mtprime}
 m_{t'} \ \; > \ \; 335\ {\rm GeV\ \ at\ 95\%\ CL},
                       \quad\quad\quad {\rm CDF},\ 4.6\ {\rm fb}^{-1},
\end{equation}
by comparing with theoretical cross sections. The expected sensitivity
is 372 GeV. Fig.~3 of CDF Note~10110 compares the seeming excess in
$H_T$ and $M_{\rm reco}$ with a $t'$ signal at 450 GeV.
The excess has less than 2$\sigma$ significance, and
if it were from a 450 GeV $t'$, the cross section would be too high.

With increased dataset, CDF has pursued another mode,
searching for $b'\bar b'$ pair production, followed by $b' \to Wt$ decay~\cite{ref:CDFbprime}.
The presence of more than 3 $W$s in the final state
allows the ultra clean signature of same-sign dileptons (much higher in
$p_T$ than the many more low $p_T$ events that lead to Eq.~(\ref{e.ASLb})),
together with missing energy and multijets.
Based on 2.7 fb$^{-1}$ data, two events are seen, one with 4 and the other with 5 jets,
whereas signal would have preferred more jets.
Because of the cleanness of the signature, a stringent
bound of~\cite{ref:CDFbprime}
\begin{equation}
 \label{e.mbprime}
 m_{b'} \ \; > \ \; 338\ {\rm GeV\ \ at\ 95\%\ CL},
               \quad\quad\quad {\rm CDF},\ 2.7\ {\rm fb}^{-1},
\end{equation}
was extracted, which should be compared with the $t'$ bound of
Eq.~(\ref{e.mtprime}). A more stringent bound can be extracted for
the so-called ``top-partner" quarks with charge $+5/3$.

While Tevatron has more data to unfold before us, the LHC has finally
started running, albeit at half the design energy at 7 TeV. It
is clear that, once the LHC has real data, it would quickly
overtake the Tevatron in the direct search of heavy new particles.
The CMS experiment has illustrated recently its potential with the
official LHC target of 1 fb$^{-1}$ at 7 TeV by the end of 2011. In
the public CMS Note-2010/008~\cite{ref:CMSn10-008}, 
CMS showed that, with just 100 pb$^{-1}$ data, 
the mass bound on $b'$ via the aforementioned same-sign dilepton 
approach would already surpass the current bound from CDF. 
These results are based on 14 TeV (done in 2008),
then 10 TeV simulation studies, then scaled down in energy to 7
TeV. The same document showed that, with 1 fb$^{-1}$ at 7 TeV, the
exclusion bound on $b'$ would reach 500 GeV. It would likely reach
beyond 500 GeV, as LO cross sections were used in the study.

Hereby one touches a new, different nerve. At 500 GeV, one is
approaching the so-called partial wave unitarity bound~\cite{ref:unitarityV},
where strong Yukawa coupling of these heavy chiral quarks
would lead to a breakdown of probability. Translated, it means
that one would need to solve the strong
Yukawa coupling theory nonperturbatively.

\subsection{Nambu Legacy --- $\overline QQ$ Condensation by Large Yukawa Coupling}

Half of the 2008 Nobel Prize went to Nambu sensei, ``for the
discovery of the mechanism of spontaneous broken symmetry in
subatomic physics".

The thought goes far back to the original observations of
Nambu in the early 1960s, in the form of the
Nambu--Jona-Lasinio (NJL) model.
In 2007, when my mind was full
of thoughts about large Yukawa couplings, nondecoupling, and CPV
for the Universe, I re-traced this thread with the help of
several papers by Bob Holdom~\cite{ref:Holdom}.
Holdom emphasized the old thought that large Yukawa coupling
(or the strong interaction theory behind~it) could lead to
$\overline QQ$ condensation, illustrating with nothing but
the old and venerable NJL model.
Discussions go back to the late 1980s, as the top grew heavier,
by Bardeen, Hill and Lindner and others, even entertaining
the thought of a heavy 4th generation.
From a ``dual" AdS/CFT strong--weak coupling correspondence point of view,
Gustavo Burdman recently promoted the ``holographic 4th generation"~\cite{ref:Burdman}.
In short, the conjecture is:
\emph{Could electroweak symmetry breaking (EWSB) be due to $b'$ and $t'$
quarks above the unitarity bound of 500--600 GeV?}
The fascination resonated with the gain of 1000 trillion (through Eq.~(\ref{e.J234}))
on CPV for heavy 4G quark masses in the range of 300--600 GeV, {\it i.e.} large Yukawa couplings.
And then Nambu received the other half of the 2008 Physics Nobel Prize.

In part because I was given the task to ``lecture on the recent Nobel prize"~\cite{ref:lectureNobel} at
FPCP2009, I literally retraced the thread that started with Nambu.
Nambu expressed strong doubts about the origin of the ``Higgs mechanism"
and the nature of the Higgs particle. In his Nobel lecture (delivered by
Jona-Lasinio), on mass generation for the gauge field, he expressed
``I thought the plasma and the Meissner effect had already established it."
In regards fermion mass generation in the electroweak Standard Model,
he compares with several examples of fermion-pairing that exhibit BCS type of SSB,
such as $^3$He superfluidity and nucleon pairing in nuclei, and comments:
``my biased opinion, there being other interpretations as to
the nature of the Higgs field", as if it should also
arise from some fermion pairing phenomena.

Let's learn once more from Nambu!

\subsection{Higgs--Yukawa on a Lattice}

Out of sheer curiosity, I started talking to local lattice theorists since
early 2008, regarding putting the Higgs--Yukawa sector on a lattice.
It seemed the natural approach towards strong Yukawa coupling, and would become
the only approach if things are nonperturbative.
Note that people are implementing ``walking technicolor" on the lattice.
For Higgs--Yukawa on a lattice, it is known that there are issues of ``triviality",
that if one sends the cutoff to infinity, the coupling constant would have to vanish.
So, strong coupling would imply that the cutoff is not far away.

Introduced through a colleague, I learned of the work of
Gerhold and Jansen, who already studied~\cite{ref:Gerhold}
the phase structure (and Higgs mass bounds) of the Higgs--Yukawa model on a lattice,
but had not explored the issues more intrinsic to EWSB.
So in the aforementioned \emph{2nd Workshop on ``Beyond 3 Generation Standard Model"}
held in January 2010, we set up a \emph{Forum} to discuss
``Higgs--Yukawa Model on a Lattice".
Through a follow-up one-day meeting in May, the intent now is to
pursue this well-defined topic.
If one could show $\langle \overline QQ\rangle \neq 0$,
i.e. $\overline QQ$ can condense through large Yukawa,
then, \emph{Who needs the Higgs for v.e.v. generation?}
Of course, in the Higgs--Yukawa model, the Yukawa coupling is
defined through the Higgs field. However, if there are two sources
for the v.e.v., then one can explore the meaning of the redundancy
of the Higgs field. Can one do away with the (elementary) Higgs field altogether?
After all, we have never observed an elementary scalar particle yet.

On a related but separate note, the study of Higgs--Yukawa on a lattice
can in principle go beyond the ``glass ceiling" of (partial wave)
unitarity violation~\cite{ref:unitarityV},
and the outcome should have implications when LHC search enters this terrain,
which could become reality with the 14 TeV run of LHC, beyond 2012.

\section{Prospects}

Measurement of $\sin2\Phi_{B_s}$ in $B_s \to J/\psi\phi$ is the current frontline, 
and recent news from CDF has weakened the discrepancy with the SM3 expectation of
$-0.04$. At the same time, the new same-sign dimuon measurement by D$\emptyset$ indicates
a 3.2$\sigma$ discrepancy with SM3 expectation. Since the value for $A_{sl}^s$ seem 
to violate a bound even with maximal $\sin2\Phi_{B_s}$, 
it seems that $\Delta \Gamma_s$ receives long-distance, hadronic corrections
 --- if the D$\emptyset$ measurement is confirmed.
If one applies the lower expectations for $\sin2\Phi_{B_s}$ from CDF, then this
``hadronic enhancement" would need to be a factor of 3 or 4.
A new physics effect on $\Gamma_{12}^s$ would have to be numerically of this order.
Unfortunately,  the same-sign dimuon asymmetry probably cannot be easily confirmed
by other experiments. Furthermore, if the trend seen by CDF is correct, then
the Tevatron would not be able to ``observe" an enhanced $\sin2\Phi_{B_s}$, 
and we have to wait for LHCb for the definitive measurement.

A similar story holds for direct search. 
Impressive limits have been extracted for both $t'$ and $b'$ by CDF. 
For $t'$ there is some nagging, unexplained activity that dampens the bound slightly.
For $b'$, an update with more data would be of great interest.
However, the Tevatron study is now approaching the limit of diminished returns,
because of dropping cross sections. At the LHC, although running at half the design energy,
once CMS and ATLAS get their data, even 100 pb$^{-1}$ would lead to bounds surpassing
the Tevatron. If 1 fb$^{-1}$ data at 7 TeV is delivered, the bound would approach 500 GeV
via $b' \to tW$ study.
Thus, LHC has good discovery potential, and in a few years, we would need to
understand what happens at or above the unitarity bound of 500 GeV or so.

If the pursuit of 4th generation quark search at the LHC bears fruit, we may 
simultaneously touch upon two of the greatest problems in particle physics, 
and even cosmology: source of EW symmetry breaking (raison d'\^etre for LHC); 
and source of CPV for BAU (raison d'\^etre for ourselves). 
There would be further implications for flavor and other physics. For example,
an early discovery of a greatly enhanced $K_L^0 \to \pi^0\bar\nu\nu$.

%\appendix

%\noindent{\bf Discussion}

%Let us go then, you and I\ldots

\acknowledgments
I thank the organizers for the invitation to speak, 
Karl Jansen for hospitality at DESY Zeuthen, 
as well as Andrzej Buras at TU Munich, 
where this writeup was finalized.

\end{document}